# Kinematical analysis of melt electrowritten jet at various print speeds

Sherry Ashour, Huaizhong Xu[*]

Department of Biobased Materials Science, Kyoto Institute of Technology, Sakyoku, Kyoto, 606-8585, Japan.

**\*Correspondence**: xhz2008@kit.ac.jp; Tel.: +81-75-724-7980

**Abstract:**

Melt electrowriting (MEW) is an extrusion-based additive manufacturing technology to create a complex construct with micro-scale fidelity. The elevated nozzle-to-collector distance of MEW needs a high requirement to control the charged jet, which relies on the understanding of the corresponding jet kinematics at different printing conditions. This study focuses on investigating the effect of printing speed on jet diameter and jet speed along the spinline from Taylor-cone to landing point, suggesting that the fiber diameter decreases linearly with the increment of printing speed; the jet speed along the spinline has a nonlinear change for printing at various speeds, i.e., the elongation rate is not constant.

1. Introduction

Melt electrowriting (MEW) is a young, high-resolution 3D printing technology that has attracted interest to make porous structures that are used for cell/tissue and biomedical materials interactions [1, 2]. MEW uses an electrohydrodynamic effect to stabilize a molten polymer jet that thins out into a small diameter (micro-scale) fiber that is direct written onto a collector [3, 4]. In addition to the high-voltage feature, MEW with an elevated nozzle-to-collector distance (1 ~ 10 mm) causes jet lag during printing, that is, the jet landing point lags behind the nozzle [5]. As the jet speed is equivalent to the collector speed, jet is perpendicular to the collector, known as critical translation speed (CTS) [6]. Printing above or at CTS is a basic principle in MEW. In addition, understanding the jet kinematics is beneficial for controlling the jet motion behaviors to achieve a stable printing process, since unreasonable print condition causes the phenomena of jet pulsing and jet break-up. Visual analysis is an efficient method to build the relationships between processing parameters and print quality. For instance, Wunner et al. developed a real-time monitoring system to identify the main effects of the processing parameters on the geometry of the jet path [7]; Mieszczanek et al. applied machine vision to analyze the volume of Taylor-cone to monitor the stability of jet during printing [8]. However, due to the ultrathin jet and high-voltage printing system, it was difficult to set a short shooting distance to measure the whole jet diameter. Recently, we established a special shooting system that can capture clear jet images [9]. Using this system, the kinematics of the jet (jet-lag angle and jet speed) at different translation speeds will be discussed for the first time.

## 2. Material and Methods

Poly(caprolactone) (PCL, $M_n$ = 45,000, Sigma-Aldrich, USA) was used as a prototype polymer since PCL can be printed for several weeks without showing obvious thermal degradation [10]. An inhouse-built MEW device was used in this study [9]. The fixing printing parameters used were a 4 kV applied voltage, 3.5 mm nozzle-to-collector distance, 75 °C melt temperature, 0.16 MPa air pressure and 23 G nozzle size, which generates a CTS of 1.3 mm/s. A camera equipped by a 300 × lens (XW200, Shenzhen Hayer Electronics Co., Ltd., China) was fixed in front of the nozzle with a working distance of 160 mm. In order to achieve clear jet images, a black plastic sheet was set behind the print head. Two lights with adjustable luminous intensity were settled close to the device to enhance the contrast of the jet image. During printing, the room temperature was controlled at 23 ± 1°C and the relative humidity was controlled at 32 ± 1%.

The jet diameter was measured by a series of image processing, including the steps of image binarization, abstraction of jet profile, smoothing, identification of spinline and segmentation (as shown in **Fig. 1c**).

The jet speed of each jet segment along the spinline was calculated by the mass conservation equation:

$$v = 4Q_v / (\text{Pi } d^2) \quad (1),$$

where $v$ is the jet speed, $Q_v$ is the volume flowrate, and $d$ is the jet diameter. The mass flow rate is 8.12 μg/s. Taking the density of 1.09 g/cm³ for the amorphous PCL melt, the

$Q_v$ is calculated to be 0.0074 mm³/s. Then the equation (1) can be simplified to

$$v = 9489.9 / d^2 \qquad (2),$$

where the unit of $v$ is mm/s, and the unit of $d$ is μm.

## 3. Results and Discussion

**Fig. 1a** displays the jet profiles taken at translation speeds of 1 ×, 1.3 ×, 1.6 ×, 1.9 × and 2.2 × CTS. With increasing the translation speed, the jet-lag angle linearly increases from 4° to 27°, and the fiber diameter linearly decreases from 92 μm to 65 μm (**Fig. 1b**); however, according to the mass conservation equation (**Equation 1**), fiber diameter ($d$) is directly proportional to $v^{0.5}$, and the jet-lag angle also should follow a nonlinear change since the angle would not exceed 90º. It should be noted that the jet-lag angle is not zero at CTS, since jet bends at the vicinity of the jet landing point, suggesting that the jet lag phenomenon cannot be fully eliminated for the MEW process. Therefore, for simplicity, within the optimum printing speed (from CTS to 2.2 × CTS), the jet-lag angle and fiber diameter can be deemed to be linear with the translation speed. According to the established database, it is convenient to achieve expected fiber diameter and jet-lag angle through changing the translation speed.

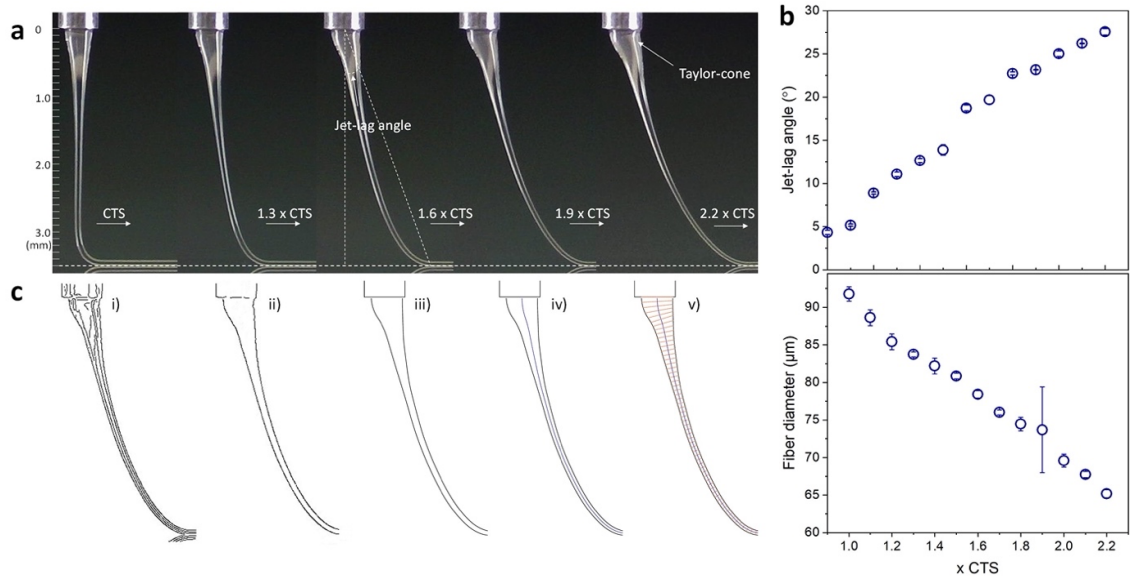

**Figure 1**. Kinematical analysis of melt electrowritten jet. (a) Jet profiles taken at 1 ×, 1.3 ×, 1.6 ×, 1.9 × and 2.2 × CTS. (b) Jet-lag angles and fiber diameters at translation speeds ranging from CTS to 2.2 × CTS. (c) Image processing: i) image binarization; ii) abstraction of jet profile; iii) smoothing; iv) identification of spinline; v) segmentation.

After image processing (**Fig. 1c**), jet diameter along the spinline was equidistantly measured, as plotted in **Fig. 2a**. It is noteworthy that at the section of Taylor-cone, the jet diameter gets larger with the increase of translation speed, which can be attributed to the deformation of Taylor-cone under drawing. At CTS, the jet diameter is relatively large as the jet is close to the collector due to jet compression, while such a phenomenon still exists at 1.3 × CTS printing speed. It can be considered as a necking phenomenon that the jet solidifies rapidly at the landing point with relatively low printing speed. When printing over 1.6 × CTS, the jet necking phenomenon disappears. Although the jet diameters along the spinline at different translation speeds have a certain deviation, the area of the whole jet (i.e., the area under the curve) is quite close, which can be regarded as a crucial parameter to evaluate the stability of the printing after changing the translation speed.

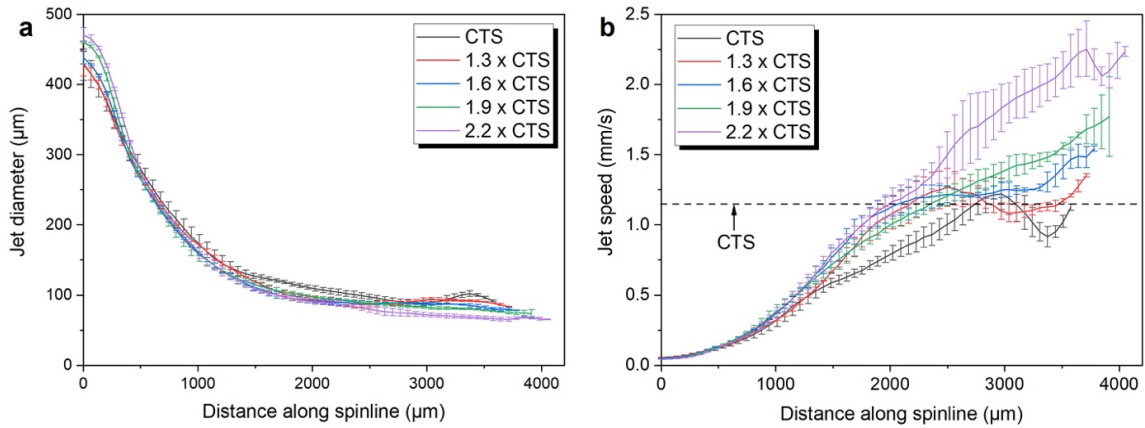

**Figure 2**. Jet diameters (a) and jet speeds (b) along the spinline.

The corresponding theoretical value of jet speed is displayed in **Fig. 2b**, calculated from the raw data of Fig. 2a according to the **Equation 2**. Since the jet sticks to the collector as landed, the jet speed at the landing point shows a speed close to the translation speed as expected. In the case of piling up at CTS, the jet speed along the spinline exhibits a linear increasing tendency for the main jet body [11]. However, there is no remarkable change of the jet speed along the spinline when printing one layer at a speed over CTS. The complex nonlinear variation indicates that the elongation rate (slope of the curve) at each jet segment is different. This is caused by the coupling effect of electrostatic force and drawing force exerted on the jet.

## 4. Conclusions

In summary, the kinematics of the MEW jet in respect of jet-lag angle, jet diameter and jet speed along the spinline at different translation speeds are investigated. Since the jet-lag angle and fiber diameter are proportional to the translation speed, it is convenient to design and print hierarchical structural scaffolds through changing the translation speed during printing. The jet speed along the spinline increases from a low speed at Taylor cone to a speed that is equivalent to the translation speed. The speed differences lead to a complex drawing process that shows nonlinear elongation rate for each jet segment along the spinline.

**Conflicts of interest:** The authors declare no competing financial interests.

**Acknowledgements**

This research did not receive any specific grant from funding agencies in the public, commercial, or not-for-profit sectors.

**References:**

[1] J.C. Kade, P.D. Dalton, Polymers for Melt Electrowriting, Adv Healthc Mater 10(1) (2021) e2001232.

[2] M. de Ruijter, A. Hrynevich, J.N. Haigh, G. Hochleitner, M. Castilho, J. Groll, J. Malda, P.D. Dalton, Out-of-Plane 3D-Printed Microfibers Improve the Shear Properties of Hydrogel Composites, Small 14(8) (2018) 1702773.

[3] K.F. Eichholz, I. Gonçalves, X. Barceló, A.S. Federici, D.A. Hoey, D.J. Kelly, How to design, develop and build a fully-integrated melt electrowriting 3D printer, Additive Manufacturing 58 (2022).

[4] F.M. Wunner, J. Maartens, O. Bas, K. Gottschalk, E.M. De-Juan-Pardo, D.W.

Hutmacher, Electrospinning writing with molten poly (ε-caprolactone) from different directions – Examining the effects of gravity, Materials Letters 216 (2018) 114-118.

[5] A. Hrynevich, I. Liashenko, P.D. Dalton, Accurate Prediction of Melt Electrowritten Laydown Patterns from Simple Geometrical Considerations, Advanced Materials Technologies 5(12) (2020) 2000772.

[6] H. Xu, I. Liashenko, A. Lucchetti, L. Du, Y. Dong, D. Zhao, J. Meng, H. Yamane, P.D. Dalton, Designing with Circular Arc Toolpaths to Increase the Complexity of Melt Electrowriting, Advanced Materials Technologies  (2022) 2101676.

[7] F.M. Wunner, P. Mieszczanek, O. Bas, S. Eggert, J. Maartens, P.D. Dalton, E.M. De-Juan-Pardo, D.W. Hutmacher, Printomics: the high-throughput analysis of printing parameters applied to melt electrowriting, Biofabrication 11(2) (2019) 025004.

[8] P. Mieszczanek, T.M. Robinson, P.D. Dalton, D.W. Hutmacher, Convergence of Machine Vision and Melt Electrowriting, Adv Mater 33(29) (2021) e2100519.

[9] S. Ashour, H. Xu, Melt electrowriting: A study of jet diameters and jet speeds along the spinline, Polymers for Advanced Technologies 33(9) (2022) 3013-3016.

[10] C. Bohm, P. Stahlhut, J. Weichhold, A. Hrynevich, J. Tessmar, P.D. Dalton, The Multiweek Thermal Stability of Medical-Grade Poly(epsilon-caprolactone) During Melt Electrowriting, Small 18(3) (2022) e2104193.

[11] S. Ashour, H. Xu, Melt electrowriting: A study of jet diameters and jet speeds along the spinline, Polymers for Advanced Technologies  (2022) 1-4.